\documentclass[twocolumn,showpacs,aps,prl,floatfix]{revtex4}

\usepackage{times,amsmath,amsfonts,amssymb,bbm,graphicx}

\newcommand{\ket}[1]{|#1\rangle}
\newcommand{\bra}[1]{\langle#1|}

\newcommand{\eq}[1]{Eq.~(\ref{#1})}

\newcommand{\beq}{\begin{eqnarray}}
\newcommand{\eeq}{\end{eqnarray}}
\newcommand{\be}{\begin{equation}}
\newcommand{\ee}{\end{equation}}
\newcommand{\ei}{\mathrm{e}}

\begin{document}

\title{Optimal estimation of losses at the ultimate quantum limit with non-Gaussian states}

\author{G. Adesso$^{1,2}$, F. Dell'Anno$^{1}$, S. De Siena$^{1}$,
F. Illuminati$^{1,3}$, and L. A. M. Souza$^{1,4}$}
\affiliation{$^{1}$Dipartimento di Matematica e Informatica,
Universit\`a degli Studi di Salerno, Via Ponte don Melillo, I-84084
Fisciano (SA), Italy; CNR-INFM Coherentia, Napoli, Italy; CNISM,
Unit\`a di Salerno; and INFN, Sezione di
Napoli - Gruppo Collegato di Salerno, Italy. \\
$^2$School of Mathematical Sciences, University of Nottingham,
University Park,  Nottingham NG7 2RD, UK.\\
$^{3}$ISI Foundation for Scientific Interchange, Viale S. Severo
65, 10133 Torino, Italy. \\
$^4$  Departamento de F\'{i}sica, Instituto de Ci\^{e}ncias Exatas,
Universidade Federal de Minas Gerais, CP 702, CEP 30161-970, Belo
Horizonte, Minas Gerais, Brasil.}

\date{March 22, 2008}

\begin{abstract}
We address the estimation of the loss parameter of a bosonic channel
probed by arbitrary signals. Unlike the optimal Gaussian probes,
which can attain the ultimate bound on precision asymptotically
either for very small or very large losses, we prove that Fock
states at any fixed photon number saturate the bound unconditionally
for any value of the loss.
In the relevant regime of
low-energy probes, we demonstrate that
superpositions of the first low-lying Fock states yield an absolute
improvement over any Gaussian probe. Such few-photon states can be
recast quite generally as truncations of de-Gaussified photon
subtracted states.
\end{abstract}

\maketitle

\noindent{\bf Introduction.---} Suppose an experimenter is given
different stations connected by channels manufactured before he/she
joined the laboratory. Assume that he/she is allowed to routinely
transmit light beams prepared e.g. in coherent, squeezed, or number
states through these channels with the purpose of implementing some
communication network. Given the state at one station, one finds out
that in general the state has been altered at the next node. The
problem is then to determine what sort of noise is affecting the
transmissions. This is a typical issue of {\em quantum parameter
estimation}, whose solution is clearly of direct
interest to practical situations like the one described. The
experimenter wishes to determine the optimal probe state that has to
be sent through the channel, and the optimal measurement that needs
to be performed at the output in order to estimate (after repeating
the process $N$ times) the loss parameter with the maximum possible
precision. In the case of amplitude damping bosonic channels, Monras
and Paris provided a solution to the problem in the particular case
of Gaussian input probe states (displaced and squeezed vacuum
states) \cite{monrasparis}. Gaussian states are easier to engineer
by quantum optical means than more sophisticated non-Gaussian
states. On the other hand, the optimal measurement needed for
loss estimation involves manipulations such as displacement and
squeezing of the output signal, and photon-counting which is a
non-Gaussian measurement not belonging to the standard toolbox of
linear optics.
At a fundamental level, using Gaussian probe inputs allows to
saturate the ultimate bound on precision  only
asymptotically in the unphysical limits of infinitesimal or infinite
losses, while in the realistic regime of intermediate loss the
Gaussian-based estimation is clearly suboptimal \cite{monrasparis}.
Therefore considering Gaussian inputs does not solve the
important problem of {\em optimal} estimation of loss in bosonic
channels, both on theoretical and practical grounds.

In this work we study the estimation of loss in bosonic channels
probed by arbitrary non-Gaussian states. For any energy of the
probes, we show that there exist non-Gaussian states improving
the estimation compared to Gaussian states in all regimes of loss.
Specifically, we prove that Fock states $\ket{n}$ (which can be
produced deterministically in the laboratory \cite{FockGen}) are the truly optimal
probes that attain the ultimate quantum limit exactly, for any $n$ and any
value of the loss. The optimal estimation then requires only
photon-counting, resulting in a technological simplification
compared to the Gaussian case. For low-energy probes (mean photon
number smaller than $1$) we construct optimal superpositions of the
first $k$ low-lying Fock states which improve the estimation over
the Gaussian case already for $k=2,3$ and approach the ultimate
limit in a much broader range of losses.
Interestingly, we find that the optimal
superpositions for $k=2$ correspond to qutrit-like, two-photon
truncations of photon-subtracted states, showing that
de-Gaussification procedures generally allow enhanced performance in
the task of loss estimation in quantum channels. This
result adds to the diverse existing instances of
non-Gaussianity as a ``power-up'' for quantum information, encountered in the optimal
cloning of coherent states \cite{Cerf}, continuous variable
teleportation \cite{CVTelepNoieKitagawa}, nonlocality tests
\cite{BellTestnonGauss}, and entanglement distillation
\cite{nonGaussDistillEnt}.

\begin{figure*}[htb]
\includegraphics[width=17cm]{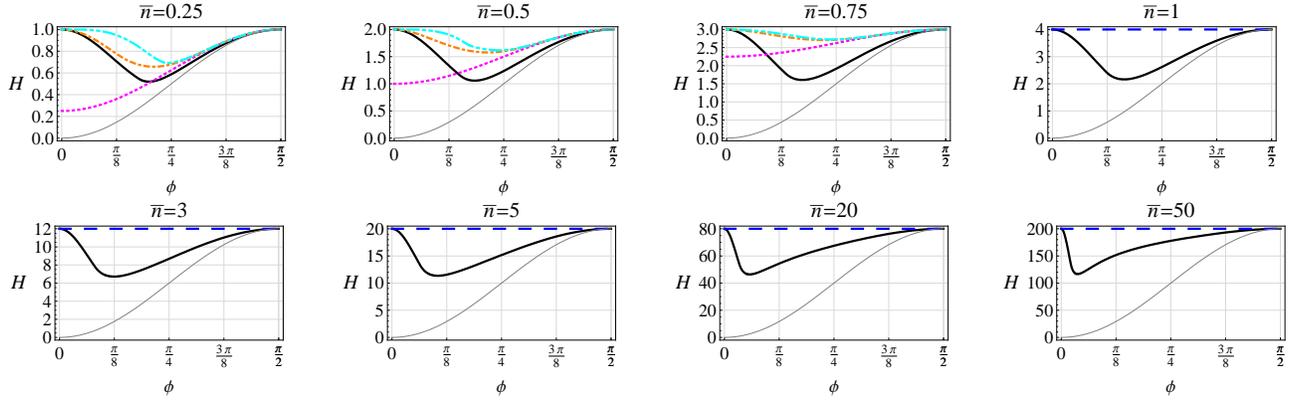}
\caption{(color online) Quantum Fisher information $H(\phi)$ versus the loss parameter $\phi$ for different values of
the input energy $\bar n$. Solid black line: optimal Gaussian
probes \cite{monrasparis}; Thin gray line: coherent states; Dashed (blue) line: Fock states
$\ket{\bar n}$; Dotted (magenta) line: qubit-like states
$\ket{\psi_0^{(1)}}$; Dot-dashed (orange) line: optimal qutrit-like
states $\ket{\psi_0^{(2)}}$; Dot-dot-dashed (cyan) line: optimal
quartet-like states $\ket{\psi_0^{(3)}}$. } \label{plhigheren}
\end{figure*}

\medskip
\noindent{\bf Bosonic channels and quantum estimation.---} We
consider a bosonic channel described by the master equation ${d
\rho}/{d t} = (\gamma/2) \mathcal{L}[a] \rho$  for quantum states
$\rho$, where $\mathcal{L}[a] \rho = 2 a \rho a^\dagger - a^\dagger
a \rho - \rho a^\dagger a$, $a$ being the annihilation operator on
the Fock space of a single bosonic mode. The aim of our study is the
optimal estimation of the loss parameter $\gamma$, or equivalently
of $\phi \in (0,\pi/2)$ defined by $\tan^2 \phi = \exp(\gamma t)-1$.
In terms of $\phi$, the master equation reads $d\rho/d \phi = \tan
\phi \mathcal{L}[a] \rho$, whose general solution is of the form \be
\label{mastersol} \rho_\phi = \sum_{n=0}^{\infty}
\frac{(\sin^2\phi)^n}{n!} (\cos\phi)^{a^\dagger a} a^n \rho_0
(a^\dagger)^n (\cos\phi)^{a^\dagger a}. \ee Let us recall the basic
elements of quantum estimation theory \cite{helstrom,holevo}, and
the relevant tools of interest for the present case
\cite{braunstein,monrasparis}. The optimal estimation of $\phi$ is
achieved asymptotically by sending $N$ independent and identically
distributed optimal probe states $\rho_0$ into the channel, and
performing at each run the optimal measurement on the output signals
$\rho_\phi$, in order to construct an estimator $\hat \phi$ to infer
the true value of $\phi$ with minimal variance. For any given
$\rho_0$, the optimal output measurement can be exactly determined
in terms of the symmetric logarithmic derivative (SLD)
$\Lambda(\phi)$, defined implicitly as the Hermitian operator that
satisfies $ {d \rho_\phi}/{d \phi} = (1/2) [\rho_\phi \Lambda(\phi)+
\Lambda(\phi)\rho_\phi]$. Using the spectral decomposition
$\rho_\phi = \sum_k \rho_k |\psi_k\rangle \langle \psi_k|$ one finds
for the SLD: \be \Lambda(\phi) = 2 \tan \phi \sum_{p q}
\frac{\langle \psi_q | \mathcal{L}[a] \rho | \psi_p
\rangle}{\rho_p+\rho_q} | \psi_q \rangle \langle \psi_p |. \ee The
resulting minimum variance saturates the quantum Cram\'{e}r-Rao
bound, $\textrm{Var}_\phi[\hat{\phi}] \geq 1/[{N H(\phi)}]$, where
the quantum Fisher information (QFI) $H(\phi)$ reads $H(\phi) =
\operatorname{Tr} [\rho_\phi \Lambda(\phi)^2]$. The problem is thus
recast in the determination of the optimal single-mode pure input
states with a given finite mean energy (or mean photon number $\bar
n$), such that the QFI of the corresponding output states is
maximal. The ultimate quantum limit on precision, that is
computable in the ideal assumption that the experimenter may have
access also to the degrees of freedom of the environment (i.e. to
the oscillators internal to the channel), is achieved for estimators
with \cite{monrasparis} $\textrm{Var}_\phi[\hat{\phi}] \geq 1/(4
{\bar n} N)$. This means that a truly optimal estimation requires
input probes which yield at the output a QFI exactly equal to $4
\bar n$ (for any single run).
If the ensemble of input signals is limited to Gaussian states
\cite{monrasparis}, the estimation is {\it never optimal}: the
ultimate limit is attained only asymptotically for $\phi$
approaching $0$ or $\pi/2$, while $H(\phi)$ for the best Gaussian
probes can get as low as $\sim 2 \bar n$ for intermediate losses
(see Fig.~\ref{plhigheren}). Here we show that non-Gaussian
probes, Fock states and low-lying superpositions thereof, are indeed
optimal for the estimation of loss in bosonic channels with the
maximum precision allowed by laws of quantum mechanics.

\begin{figure*}[htb]
\includegraphics[width=17cm]{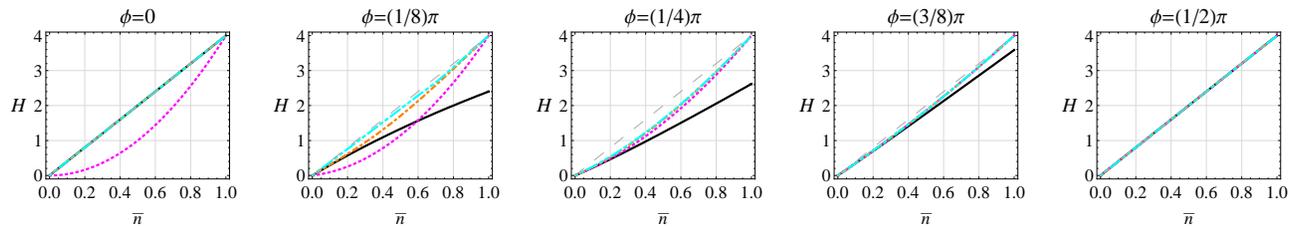}
\caption{(color online) Quantum Fisher information $H(\phi)$ versus the input energy $\bar n$ for different values of
the loss parameter $\phi$. Solid line: optimal Gaussian probes
\cite{monrasparis}; Dotted (magenta) line: qubit-like states
$\ket{\psi_0^{(1)}}$; Dot-dashed (orange) line: optimal qutrit-like
states $\ket{\psi_0^{(2)}}$; Dot-dot-dashed (cyan) line: optimal
quartet-like states $\ket{\psi_0^{(3)}}$. The thin dashed line
depicts the ultimate quantum limit, $H(\phi)=4 \bar n$.}
\label{figlowen}
\end{figure*}

\medskip
{\noindent \bf Fock states.---} Let us consider, as input probes,
Fock states $\rho_0 = |n\rangle \langle n|$ ($\bar n = n$).  The evolved state, according to
\eq{mastersol}, reads $\rho_\phi = \sum_{k=0}^{n}(\sin^2\phi)^k
\binom{n}{k} (\cos^2\phi)^{(n-k)} |n-k \rangle \langle n-k |.$ The
SLD for this case is $\Lambda(\phi) = \tan \phi \sum_{k=0}^{n}
({g_k}/{f_{n-k}}) |k\rangle \langle k|$, where $g_k = 2 \left[
f_{n-k-1} (k+1) (1-\delta_{k,n}) - f_{n-k} k \right]$ and $f_{k} =
\binom{n}{k} (\sin^2 \phi)^k (\cos^2 \phi)^{n-k}$. The QFI reads
$H(\phi) = \tan^2 \phi \sum_{k=0}^{n} (g_k^2/f_{n-k}) = 4n.$ Fock
states thus enable the {\em optimal unconditional} estimation of
loss regardless of the actual value of the parameter to be estimated
(see Fig.~\ref{plhigheren}). This makes an adaptive estimation
scheme unnecessary (unlike the Gaussian case \cite{monrasparis}).
Moreover, the measurement that has to be performed, obtained by
projecting onto one-dimensional eigenspaces of $\Lambda(\phi)$
\cite{braunstein}, can be implemented only by simple
photon-counting. Given the recently achieved degree of control
in this measuring technique \cite{Haroccia} and in
the high-fidelity engineering of Fock states with a small number $n \lesssim 10$ of photons (conditionally for running optical fields, and even
deterministically in microwave cavity or circuit QED) \cite{FockGen}, with $n=2$ standing as an ideal workpoint, our results might pave the way for an experimental
verification of the quantum theory of optimal estimation and a
measurement of the SLD to infer the value of such a relevant
parameter as the loss factor in dissipative channels.
While one may argue that in practice it is easier to produce and manipulate  ``classical'' fields, i.e. coherent states obtained from attenuated laser beams, than nonclassical resources such as squeezed (Gaussian) and Fock states, we remark that, as shown in Fig.~\ref{plhigheren}, the performance of coherent probes for loss estimation is quite far from optimality. In particular in the regime of small and intermediate
losses ($\phi < \pi/4$), corresponding to routinely available good quality fibre channels, the precision achieved by, say, a two-photon Fock state would be matched by that of a coherent field with much higher mean photon number (e.g. ${\bar n}=26$ for $\phi=\pi/16$, and ${\bar n}=105$ for $\phi=\pi/32$), an increase in energy which may  be not worth paying in terms of efficiency of the estimation. In
fact, in actual implementations it is very desirable to have probes
of low energy in order not to alter the channel significantly \cite{parisuniversal} and to
enable repeatability of the input-and-measure scheme.
In this respect it appears crucial to  identify classes of non-Gaussian states which
may attain the ultimate precision for any value of the input energy, especially in the relevant regime of $0 \le \bar n \le 1$.

\medskip
{\noindent \bf Photonic qubit states.---} The simplest candidate
probe state in the low-energy regime is the superposition of the
vacuum and the one-photon Fock state,
$\rho_0=\ket{\psi_0^{(1)}}\!\bra{\psi_0^{(1)}}$, $\ket{\psi_0^{(1)}}
= \cos\theta \ket0 + \ei^{i \varphi} \sin\theta \ket1$,
characterized by a mean photon number $\bar n = \sin^2\theta$. It is
rather straightforward to obtain the evolved state $\rho_\phi$ and
to diagonalize it in order to compute the SLD. The resulting
expression for the QFI is found to be independent of the phase
$\varphi$ and given by $H^{(1)}(\phi) = 4 {\bar n} [1-(1-{\bar n})
\cos^2\phi]$. We notice (see Figs.~\ref{plhigheren}
and~\ref{figlowen}) that  the considered simple example of
non-Gaussian superposition state (with no free parameter left for
optimization) yields a significant improvement over the best
Gaussian estimation for intermediate-high losses, although in the
regime of small losses and small energies Gaussian states (which in
this limit are simply squeezed states \cite{monrasparis}) remain
better probes.

\begin{figure}[bt]
\includegraphics[width=7cm]{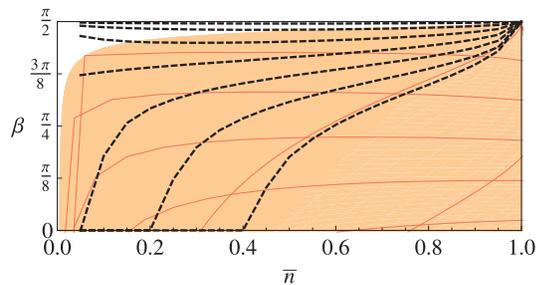}
\caption{(color online) Shaded surface: attainable values in the
space of the coefficients  $\beta$ and $\bar n$ of qutrit-like
states, as functions of the parameters $r$ and $\eta$ associated to
the truncation of photon-subtracted states. Dashed lines: optimal
$\beta$ as a function of $\bar n$, corresponding to the maximum
quantum Fisher information among all qutrit states, for different
values of the loss parameter $\phi$ (ranging from $\pi/16$ to
$\pi/2$ from bottom to top).} \label{figatronca}
\end{figure}

\medskip
{\noindent \bf Photonic qutrit states.---} Next, we consider
superpositions of the vacuum and the first two Fock states,
$\rho_0=\ket{\psi_0^{(2)}}\!\bra{\psi_0^{(2)}}$, with
$\ket{\psi_0^{(2)}} =\cos\alpha \ket0 + \ei^{i \mu} \sin \alpha \sin
\beta \ket1 + \ei^{i \nu} \sin\alpha \cos\beta \ket2$. Here $\alpha$
can be fixed as a function of $\beta$ and $\bar n$, $\alpha={\rm
arcsin}\left(\sqrt{\frac{2 \bar n}{\cos (2 \beta )+3}}\right)$. The
evaluation of the SLD involves the  diagonalization of the $3 \times
3$ matrix corresponding to the output state. The QFI has to be
optimized, for a given $\bar n$, over the phases $\mu$ and $\nu$ and
over the weight $\beta$ (the latter ranges from $\beta=0$,
corresponding to a superposition of $\ket0$ and $\ket2$, to
$\beta=\pi/2$, corresponding to the previously considered qubit-like
state superposition of $\ket0$ and $\ket1$). Maximization over the
phases yields $\mu=\nu=\pi$. The optimal $\beta$ can instead be
found numerically for each $\bar n$, $\phi$, and is reported in
Fig.~\ref{figatronca}. The resulting optimal QFI is shown in
Figs.~\ref{plhigheren} and~\ref{figlowen}. The photonic qutrit
states improve over the qubit-like state and, more remarkably, over
the optimal Gaussian probes with the same mean energy in the whole range of parameters (i.e., for any value of the loss). In the limit of vanishing probe energy, $\bar
n \rightarrow 0$, the optimal Gaussian probe is a purely squeezed
vacuum \cite{monrasparis} with QFI $H^{(G)}(\phi)=4 {\bar n} [1+
z^2]/[1+2z(1+{\bar n})+z^2]$ (where $z=\tan^2\phi$), while the
optimal qutrit-like state is a pure superposition of $\ket0$ and
$\ket2$ ($\beta=0$) with QFI $H^{(2)}(\phi) = 4 \bar n
[1+z^2]/[1+z(2-{\bar n} + z] \ge H^{(G)}(\phi)$. In the limit $\bar
n \rightarrow 1$, the weight $\beta$ increases up to $\pi/2$ and the
qutrit-like state converges to the optimal Fock state $\ket1$.

\medskip
{\noindent \bf Relation to de-Gaussified states.---} A very
important and natural question concerns the nature of such an
optimal qutrit-like state, in particular whether such state can be
interpreted as a finite-dimensional truncation of an
infinite-dimensional non-Gaussian state, and how to determine the
latter. To this aim, we consider de-Gaussified photon-subtracted,
displaced, and squeezed states $\rho_{0}^{(nG)} = \mathcal{N}^{-1}
a D(\eta) S(r) |0\rangle \langle 0| S^\dagger(r) D^\dagger (\eta)
a^\dagger$ and study their projections on the subspace spanned by
the vacuum and the first two Fock states. The choice for comparison
is inspired by the fact that the strategies of photon
addition and subtraction are the current royal avenues to the
experimental production of optical non-Gaussian resources \cite{nonGaussGen}. Truncation of the photon-subtracted
displaced squeezed state $\rho_{0}^{(nG)}$ yields the state
$|\psi^{(nGtr)}\rangle= c_0 \ket0 + c_1 \ket1 + c_2 \ket2$, where
the coefficients $c_j=k_j/(k_0^2+k_1^2+k_2^2)^{1/2}$, with $k_0 =
\eta  (\tanh r+1)$, $k_1=k_0^2-\tanh r$, $k_2=2^{-1/2}k_0(k_0^2-3
\tanh r)$. The coefficients $k_{j}$ are functions of the modulus of
the displacement $\eta$ and the real squeezing amplitude $r$ (the
relative phase can be set to zero in order to maximize the QFI).
Remarkably, we find that for any $\eta$ and $r$ such that $0\leq
\bar{n} \leq 1$, the pure states $\rho_{0}^{(nG)}$ and
$\rho_{0}^{(nGtr)}=|\psi^{(nGtr)}\rangle \langle\psi^{(nGtr)}|$
always possess a high-fidelity overlap
$\mathcal{F}={\rm Tr}[\rho_{0}^{(nG)}\rho_{0}^{(nGtr)}]>92\%$. Adding one
or few further terms in the superpositions quickly rises the
fidelity well above $99\%$. The mean photon number $\bar n$ is a
highly nonlinear function of $\eta$ and $r$. Thus, in order to
visualize the maximization of $H$ over these two resource
parameters, at fixed $\bar n$, we let $\eta$ and $r$ vary in the
real space, and depict as a shaded region in Fig.~\ref{figatronca}
the achievable range of $\beta$ (characterizing the originally
introduced parametrization of fixed-energy qutrit states) versus the
values of $\bar n$ that are spanned by the variation of the
parameters. We find that there exist combinations of values of
$\eta$ and $r$ for the truncated photon-subtracted states such that
the angle $\beta$ can take almost any value in the range $[0,\pi/2]$
for each $\bar n \in [0,1]$. Superimposing the parametric region
with the curves of the optimal $\beta$ as a function of $\bar n$
yields the maximum QFI among all qutrit states, for different values
of the loss parameter $\phi$. Apart from a small range of extremely
high losses and low energies (in which practically all states, such
as non-truncated Gaussian states, qubit states, qutrit states, etc..
yield the same optimal performance close to the ultimate limit),
there always exist values of the displacement $\eta$ and of the
squeezing $r$ such that the truncated photon-subtracted state
reproduces exactly the optimal qutrit state (pictorially, the dashed lines fall in the
attainable shaded region in Fig.~\ref{figatronca}). On the other
hand, this conclusion does not apply to finite truncations of
Gaussian states, which are never optimal among all pure qutrit
states in many ranges of values of energies and losses, although
they can still perform better than the original Gaussian states.
This clearly shows that de-Gaussification, be it implemented by
means of truncation, of photon-subtraction, or both, generally
enhances the task of estimating loss in bosonic Gaussian
channels.
It is reasonable
to conjecture that there exist particular families of non-Gaussian
 states (e.g. non-truncated photon-subtracted states) that represent optimal resources
for the considered task in the regime of low energy, attaining the ultimate quantum limit also for intermediate
losses, where superpositions of the first low-lying Fock states
do not saturate the $4 \bar n$ scaling of $H(\phi)$.

\medskip
{\noindent \bf Higher-order superpositions.---} The previous
conclusions can be confronted by investigating the effect of adding
terms of higher order in the superpositions. Consider states of the
form $\rho_0=\ket{\psi_0^{(3)}}\!\bra{\psi_0^{(3)}}$ with
$\ket{\psi_0^{(3)}}=\sum_{n=0}^3 c_n \ket{n}$, i.e. superpositions
of Fock states up to $n=3$. The optimal QFI can be obtained by
optimizing numerically the complex weights $c_n$ for each $\bar n$,
$\phi$ (see Figs.~\ref{plhigheren} and~\ref{figlowen}). This yields
a further improvement over the optimal qubit and qutrit states as
well as over the Gaussian states. The succession of curves of
$H^{(k)}(\phi)$ thus appears to converge to $4 \bar n$ for $k
\rightarrow \infty$. It is certainly true that the best possible performance requires
non-Gaussian states in every energy range and for any value of the
loss. Obviously, not all non-Gaussian states
improve over Gaussian ones: For instance,  the QFI for cat-like  superpositions of coherent states   is almost always smaller than
that of the optimal Gaussian probes, but for  $\bar n \lesssim 2$ and $\phi \lesssim \pi/8$.

\medskip
{\noindent \bf Discussion.---} Estimating the loss factor of a
channel is an important theoretical issue of direct practical
relevance. In the case of purely dissipative bosonic channels, we have
shown that proper nonclassical non-Gaussian states such as Fock states and superpositions thereof need to be
employed as probes in order to achieve the most precise estimation. The optimality of Fock states can be understood on intuitive grounds by realizing that the parameter subject to estimation is essentially the decay rate of the field energy, and Fock states are eigenstates of the energy observable thus having zero energy uncertainty in their preparation.

The present
work provides strong support for the need of going beyond the
Gaussian scenario in applied quantum  technology and quantum metrology, and
motivates further research to realize advanced tools of non-Gaussian
quantum state engineering, manipulation  and detection.
\\ \quad \\ \noindent L.A.M. Souza was supported by CNPq-Brasil.


\end{document}